\documentclass[5p]{elsarticle}

\usepackage{hyperref}
\usepackage{graphicx}
\usepackage{amssymb}
\usepackage{amsmath}
\usepackage{lineno}
\usepackage{color}

\biboptions{square,comma,sort&compress}

\journal{Computer Physics Communications}

\begin{document}

\begin{frontmatter}

\title{Adsorption of polymers at nanowires}

\author{Thomas Vogel}
\ead{thomasvogel@physast.uga.edu}
\author{Michael Bachmann}
\ead{m.bachmann@fz-juelich.de}
\ead[url]{http://www.smsyslab.org}

\address{Soft Matter Systems Research Group,\\ Institut f\"ur Festk\"orperforschung (IFF-2) and Institute for Advanced Simulation (IAS-2),\\ Forschungszentrum J\"ulich, D-52428 J\"ulich, Germany}

\begin{abstract}
Low-energy structures of a~hybrid system consisting of a~polymer and
an attractive nanowire substrate as well as the thermodynamics of the
adsorption transition are studied by means of Monte Carlo computer
simulations. Depending on structural and energetic properties of the
substrate, we find different adsorbed polymer conformations, amongst
which are spherical droplets attached to the wire and monolayer tubes
surrounding it. We identify adsorption temperatures and the type of
the transition between adsorbed and desorbed structures depending on
the substrate attraction strength.
\end{abstract}


\end{frontmatter}

\section{Introduction}
\label{intro}

In a~systematic study of low-energy states of polymers adsorbed to a~stringlike substrate, we recently found a~variety of different
conformational phases by just changing two basic substrate
properties~\cite{tv10prl}. Among these structures are spherical
droplets attached to the string as well as barrellike monolayer
conformations surrounding the string. The latter conformations exhibit
similarities to single-walled
carbon nanotubes.

Generally, the adsorption of polymers on material surfaces or
substrates is a~crucial and nontrivial process in nature and
nanotechnology. It is also known, for example, that the adsorption
process or, in fact, the potential of a~polymer to adsorb at
a~semiconductor surface depends essentially on details like the exact
position of single monomers in the primary structure of
a~hetero\-polymer~\cite{bachmann10acie}. However, specific hybrid
systems composed of inorganic matter and polymers potentially
facilitate the development of completely new nanotechnological devices
like sensors for specific single molecules or devices for ultrafast
photonics~\cite{gao03ea,hasan09am}.

Hence, fundamental investigations are inevitable for a~better understanding of such systems. Computational studies of
coarse-grained systems have proven to be adequate and quite useful
for this purpose in recent years, both for predicting and
interpreting specific and basic behavior of polymer
adsorption~\cite{milchev01jcp,bachmann05prl,bachmann06pre1,monika09jpcb,bachmann10acie}.

Particularly interesting are systems with a~cylindrical substrate,
like carbon nanotubes. Many of the special properties of these
structures, that make them potential candidates for technological applications,
can be controlled, influenced or amplified by coating them
with polymeric material~\cite{gao03ea,hasan09am}. In previous
computational works, for example, the wetting of cylindrical fibers
with polymers or the helical wrapping of single polymers around
nanocylinders were studied~\cite{milchev02jcp,srebnik07cpl}.

In a~recent study, we have revealed a~more general
picture of the adsorption of polymers at different ultrathin
cylindrical substrates~\cite{tv10prl} by performing generalized-ensemble
Monte Carlo simulations~\cite{bergneuh91plb,bergneuh92prl,wangl01prl}
for a~general coarse-grained model of this hybrid system. Here, after
describing technical details (Sects.~\ref{model} and~\ref{meths}), we focus on
specific conformational transitions at very low temperatures
(Sect.~\ref{gs}), and discuss in a~thermodynamic analysis the adsorption
transition at comparatively high temperatures (Sect.~\ref{thermo}).

\section{The model}
\label{model}

For our adsorption study, we employ a~coarse-grained off-lattice model, where
the polymer consists of identical monomers which are represented by beads
without internal structure. These are connected sequentially by stiff bonds
of unity length. In order to facilitate future enhancements and the
comparison with previous studies (see, e.g.,~\cite{monika09jpcb}), we
introduce a~weak stiffness between the bonds, i.e., the polymer is not
strictly flexible. The polymer is placed into a~simulation box which also
contains an attractive thin string located in its center. Its orientation defines the $z$-axis.
The edge
lengths of this box in $x$ and $y$ directions with 
periodic boundary conditions are chosen to be twice as large as the length of the completely
stretched polymer. We note 
that the polymer is not grafted to the string and
may move freely in space.

The total energy $E$ of the polymer consists of contributions from the
Lennard-Jones interaction $V_\mathrm{LJ}$ between all pairs of nonadjacent
monomers, a~weak
bending stiffness $V_\mathrm{bend}$ and the monomer--string interaction
$V_\mathrm{string}$:
\begin{equation}
\begin{split}
E=&\sum_{i=1,j>i+1}^{N-2}V_\mathrm{LJ}(r_{ij})+\sum_{i=2}^{N-1}V_\mathrm{bend}(\cos\theta_i)\\
&+\sum_{i=1}^{N}V_\mathrm{string}(r_{\mathrm{z};i})\,,
\end{split}
\end{equation}
with
\begin{align}
&V_\mathrm{LJ}(r_{ij})=4\epsilon_\mathrm{m}\left[\left(\frac{\sigma_\mathrm{m}}{r_{ij}}\right)^{12}-\left(\frac{\sigma_\mathrm{m}}{r_{ij}}\right)^{6}\right],\tag{1a}\label{eq:1_lj}\\
&V_\mathrm{bend}(\cos\theta_i)=\kappa\,(1-\cos\theta_i)\,,\tag{1b}\\
&V_\mathrm{string}(r_{\mathrm{z};i})=\pi\,\eta_\mathrm{f}\epsilon_\mathrm{f}\left(\frac{63\,\sigma_\mathrm{f}^{12}}{64\,r_{\mathrm{z};i}^{11}}-\frac{3\,\sigma_\mathrm{f}^{6}}{2\,r_{\mathrm{z};i}^5}\right),\tag{1c}\label{eq:1_str}
\end{align}
where $r_{ij}$ is the geometrical distance between two mono\-mers $i$
and $j$, $\theta_i$ is the angle between the two bonds connected to
monomer $i$, and $r_{\mathrm{z};i}$ is the distance of the $i$th
monomer perpendicular to the string. The
parameters are set as follows:
$\epsilon_\mathrm{m}=\sigma_\mathrm{m}=1$, such that
$V_\mathrm{LJ}(2^{1/6})=-1$. The bending
stiffness is chosen to be comparatively weak, $\kappa=0.25$~\cite{stefan07prl,monika09jpcb}.

The interaction $V_\mathrm{string}$ between the string and the
mono\-mers is also based on a~simple Lennard-Jones potential, where
the wire is assumed to have a~homogeneous ``charge''
distribution~\cite{milchev02jcp,srebnik07cpl,monika09jpcb}. 
The string potential can then be considered as the
limiting case of the potential of a~cylinder~\cite{milchev02jcp} in
the limit of vanishing radius and keeping the overall charge
fixed~\cite{tvlong10tbp}.

\begin{figure}[b!]
\includegraphics[width=\columnwidth]{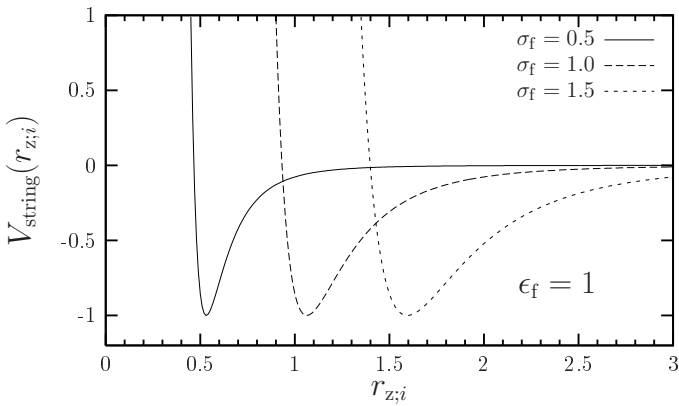}
\caption{The interaction potential between a~monomer and the
  string. Note that by scaling the string ``charge'' density with
  $\sigma_{\mathrm{f}}^{-1}$, the minimum value of the potential is
  $-1$ independently of $\sigma_{\mathrm{f}}$.}
\label{fig1}
\end{figure}

Alternatively, the Lennard-Jones potential for the interaction between
a~monomer and the string can be integrated out along the string axis
to yield~(\ref{eq:1_str}),
\begin{equation}
  V_\mathrm{string}(r_{\mathrm{z}})=4\,\eta_\mathrm{f}\epsilon_\mathrm{f}\int_{-\infty}^{\infty}\mathrm{d}z
  \left[\frac{\sigma_\mathrm{f}^{12}}{(r_{\mathrm{z}}^2+z^2)^{6}}-\frac{3\,\sigma_\mathrm{f}^{6}}{(r_{\mathrm{z}}^2+z^2)^{3}}\right]\,,
\end{equation}
where $\sigma_\mathrm{f}$ is the van der Waals radius of the string and
can be considered as its effective ``thickness''. It is related
to the minimum distance $r_{\mathrm{z}}^\mathrm{min}$ of the string
potential via
\begin{equation}
\label{eq:r_vs_sigma}
r_{\mathrm{z}}^\mathrm{min}(\sigma_\mathrm{f})=\left(\frac{693}{480}\right)^{1/6}\sigma_\mathrm{f}\approx1.06\,\sigma_\mathrm{f}\,.
\end{equation}
For convenience, we scale the string ``charge'' density
$\eta_\mathrm{f}$ in such a~way that the minimum value of the
potential is $V_\mathrm{string}(r_{\mathrm{z}}^\mathrm{min})=-1$
independently of $\sigma_\mathrm{f}$, i.e., we set $\eta_{\rm
  f}\approx 0.53\,\sigma_f^{-1}$.  Fig.~\ref{fig1} shows the
correspondingly scaled string potential for different values of
$\sigma_\mathrm{f}$.

\section{Simulational details}
\label{meths}

Polymer systems are known to possess highly nontrivial, rugged
free-energy landscapes~\cite{janke08buch}. For the simulation, we
have, therefore, applied generalized-ensemble Monte Carlo methods. The
ground-state energies have been estimated by using various, but
conceptually similar, stochastic methods such as parallel tempering,
Wang--Landau, and multicanonical
sampling~\cite{bergneuh91plb,bergneuh92prl,janke98physA,wangl01prl},
as well as especially designed optimizing approaches like energy
landscape paving, where the energy landscape is deformed irreversibly
during the simulation~\cite{wenzel99prl,hansm99epjb,hansm02prl}. The efficiency of all
stochastic methods strongly depends on the conformational update set
used (and, of course, on the fine-tuning of each method). If a~move
set is chosen reasonably well, all methods lead to comparable results
in similar times.  In addition, we have refined low-energy states by
standard deterministic optimization techniques such as the conjugate
gradient method.

For the estimation of the density of states and hence all
thermodynamic quantities, we employed the Wang-Landau
method~\cite{wangl01prl,zhoubhatt05pre} for the determination of the
multicanonical weights and performed a~final production run in the
multicanonical
ensemble~\cite{bergneuh91plb,bergneuh92prl,janke98physA}. Independently
of the values of the potential parameters, we partition the simulated
energy interval into 10\,000 bins in each simulation. The actual bin
size hence depends on the energy range delimited by the putative
ground-state energy and a~fixed boundary on high energies.

\begin{figure}[b!]
\includegraphics{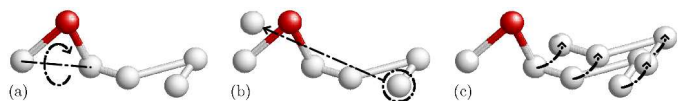}
\caption{Conformational update moves used in our
  simulations. (a)~local crankshaft move, (b) slithering snake or
  reptation move, and (c) global spherical cap update.}
\label{fig2}
\end{figure}

As known and mentioned above, the choice of the update scheme is
crucial for the efficiency of any Monte Carlo simulation in
general. For conformational changes, we apply here a~variety of update
moves, see Fig.~\ref{fig2}, including local crankshaft
[Fig.~\ref{fig2}(a)] and slithering-snake moves [Fig.~\ref{fig2}(b)],
as well as global spherical-cap [Fig.~\ref{fig2}(c)] and translation
moves. Sets including these steps have been found to work quite well
in previous studies as
\hbox{well~\cite{schnabel09cpl,schnabel09jcp,taylor09jcp}}. The crankshaft
move is just a~rotation of a~single monomer around the virtual bond
between its neighbors, or, if the monomer is an end-monomer, around
the elongation of the one neighboring bond. For the slithering-snake
update, we cut a~monomer from one end and paste it at the other end
keeping the bond vector fixed. Both updates induce small
conformational changes of the whole chain, whereas the latter one
enables the polymer to leave very dense, adsorbed conformations.

The spherical-cap update consists of the shift of $1\leq n<N$ monomers by
a~small constant vector keeping the bond lengths at the $n$th monomer
fixed. It hence allows for larger steps in the conformational space
compared to the former local updates. The global translation update
finally allows for a~direct displacement of the chain relative to the
string, which in the entire simulation remains fixed in the box.  In
any Monte Carlo step, we choose the different moves randomly with
equal weight. We convinced ourselves that this provides a~reasonable
sampling of the conformational space of this model system on large
scales as well as locally. It behaves in general not worse and for the
ground-state search even better than a~procedure consisting of a~few
global moves followed by much more local moves, whereas the ratio
depends on the system size. However, such a procedure has been found to be
favorable for other problems~\cite{taylor09jcp}.

Alternative, more sophisticated update
moves like bond-bridging or monomer-jump
moves~\cite{deutsch97jcp,schnabel09jcp,taylor09jcp,reith10cpc} have
not been included in our update set as they are rather time consuming
and would apparently not improve the principal findings of the present
study. Though, they are necessary for studying structural transitions
in the dense and crystalline
regime~\cite{schnabel09cpl,schnabel09jcp}, or for the investigation of
much larger systems. In the following, we discuss the structural
properties of a~polymer with $N=100$ monomers.

\section{Ground states of the system}
\label{gs}

\begin{figure}
\includegraphics{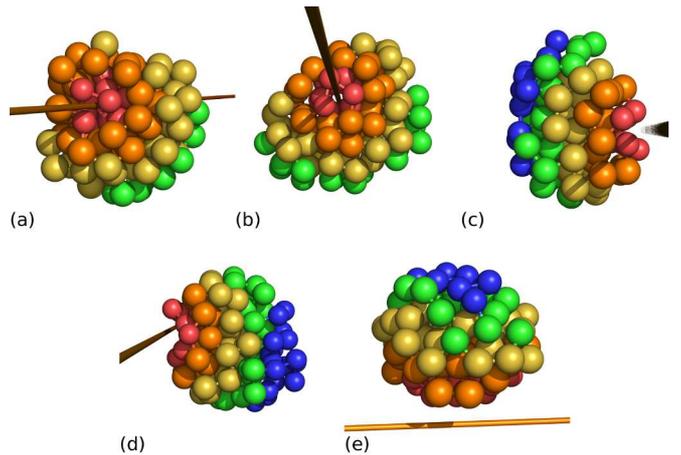}
\caption{Low-energy conformations for $\epsilon_{\mathrm{f}}=1$ and (a)
  $\sigma_{\mathrm{f}}=0.55$, (b) $0.573$, (c) $0.647$, (d) $1.0$, and (e)
  $1.5$. (a) and (b) correspond to phase Gi, conformations in (c)--(e)
  belong to phase Ge.}\vskip-.2\baselineskip
\label{fig3}
\end{figure}

There are two parameters in the interaction potential between the
string and monomers [Eq.~(\ref{eq:1_str}) and Fig.~\ref{fig1}], the
van der Waals radius proportional to the effective thickness of the
string, $\sigma_{\mathrm{f}}$, and the adsorption strength
$\epsilon_{\mathrm{f}}$. By varying these two parameters, we have
recently constructed the complete conformational phase diagram of
low-energy structures~\cite{tv10prl,tv10procathens}. By introducing
suitable observables, four major structural phases have been
identified. For small values of $\epsilon_{\mathrm{f}}$ and
$\sigma_{\mathrm{f}}$, i.e., for very thin and weakly attracting
strings, we find globular or spherical polymer droplets inclosing the
string (phase Gi). In this case, the polymer structures are similar to
that in bulk under poor solvent conditions. The string does not
influence the shape of these structures but affects only the internal
structure of the droplet. 

Figure~\ref{fig3} shows conformations with
$\epsilon_{\mathrm{f}}=\epsilon_{\mathrm{m}}=1$. In
Figs.~\ref{fig3}(a) and~\ref{fig3}(b), droplets inclosing the string
are visualized. When increasing the van der Waals radius of the
string, monomer--monomer bonds inside the droplet will be broken and,
hence, the string is excluded from the droplet but still it is
attached to it (phase Ge). The radius at which this rearrangement
occurs depends on $\epsilon_{\mathrm{f}}$. For
$\epsilon_{\mathrm{f}}\lesssim\epsilon_{\mathrm{m}}$, i.e., where the
string attraction is not significantly stronger than the
monomer--monomer attraction, the transition occurs, roughly, when the
diameter of the string becomes comparable to the equilibrium distance
between two monomers. When further increasing the string radius, the
string moves outward and the structure approaches the bulk
conformation. See Figs.~\ref{fig3}(c)--(e) for examples of
conformations at different $\sigma_\mathrm{f}$ values, thus
visualizing the described ``process''.

\begin{figure*}[t]
\includegraphics{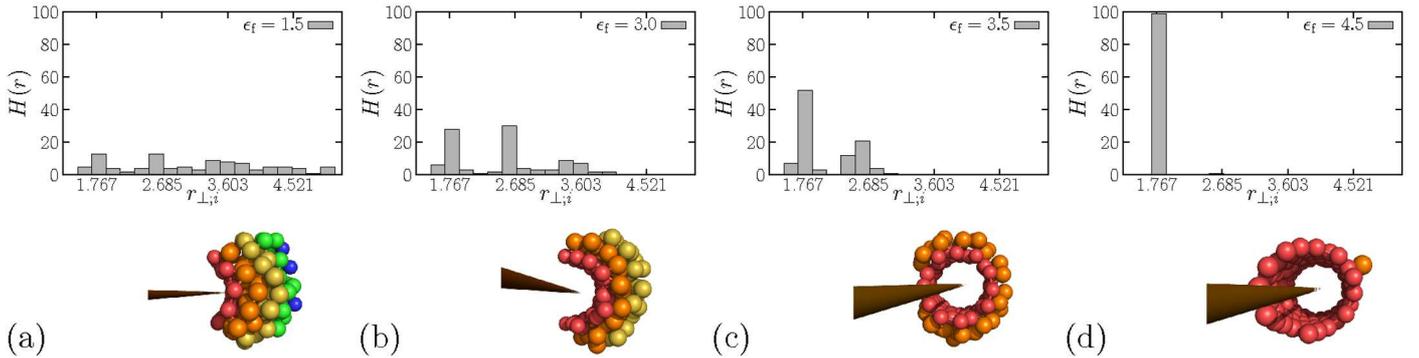}
\caption{Top: Radial distribution functions of low-energy states with
  $\sigma_{\mathrm{f}}=5/3$ and (a) $\epsilon_{\mathrm{f}}=1.5$, (b)
  $3.0$, (c) $3.5$, and (d) $4.5$. Bottom: Visualizations of the
  respective structures. Different colors or shapes encode different
  monomer layers, i.e., regions within certain distances to the string.}
\label{fig4}
\end{figure*}

Increasing at any given value of $\sigma_{\mathrm{f}}$ the string attraction
strength $\epsilon_{\mathrm{f}}$, globular structures (Gi, Ge) start
to deform and to lose spherical symmetry at
$\epsilon_{\mathrm{f}}\gtrsim3$. For $\sigma_{\mathrm{f}}\lesssim1.5$,
we observe a~transition from phase Gi directly to the barrel phase B,
which is characterized by closed, stretched conformations with 
cylinderlike shape wrapping around the string. In the extreme case of
very strong string attraction, polymers form monolayer tubes. 
For $\sigma_{\mathrm{f}}\gtrsim1.5$, the polymer first
adopts clamshell-like conformations (phase~C) before it forms barrel
structures in phase B. Interestingly, the evolution from spherical
droplets to monolayer tubes involves the formation of distinguishable
monomer layers. 

For illustration, we plot in the upper row in Fig.~\ref{fig4} the
radial distribution of monomers with respect to the string of certain
low-energy structures. One finds accumulations of monomers at
different distances from the string, i.e., in different layers. The
position of the first layer corresponds to the van der Waals radius of
the string [cp.\ Eq.~(\ref{eq:r_vs_sigma}),
  $\sigma_{\mathrm{f}}=5/3$], whereas the location of the higher-order
layers is connected to the equilibrium distance between the monomers,
corresponding to $\sigma_{\mathrm{m}}$. In the lower row, respective
structures are depicted. In Fig.~\ref{fig4}(a), we plot the radial
distribution of monomers in a~conformation from phase Ge with
$\epsilon_{\mathrm{f}}=1.5$. The emergence of different peaks in that
function can be observed. A~clear 3-layer structure can be identified
in Fig.~\ref{fig4}(b), where a~typical conformation in phase C is
shown ($\epsilon_{\mathrm{f}}=3$). In Fig.~\ref{fig4}(c), a~two-layer
barrel-shaped conformation is shown ($\epsilon_{\mathrm{f}}=3.5$)
which transforms into a~monolayer tube at
$\epsilon_{\mathrm{f}}\gtrsim4.5$, as depicted in
Fig.~\ref{fig4}(d). We would like to note here that, in particular,
these monolayer tubes exhibit interesting similarities to other
structures in nature like, for example, carbon
nanotubes~\cite{tv10prl,tv10procathens,tvmbtmja10tbp1,tvmbtmja10tbp2}.

\section{Thermodynamics of the adsorption}
\label{thermo}

\begin{figure}[t!]
\includegraphics{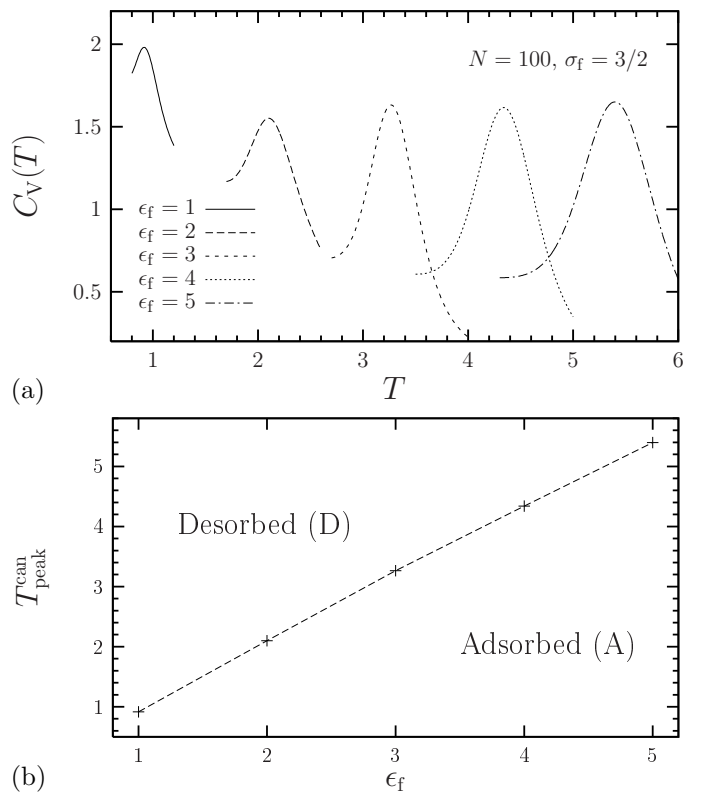}
\caption{(a) Adsorption peaks in the canonical heat capacities of the
  100-mer for different values of $\epsilon_{\mathrm{f}}$ at
  $\sigma_{\mathrm{f}}=3/2$ [cp.\ the peak temperatures with
  temperatures from the microcanonical analysis in
  Fig.~\ref{fig6}(b)]. (b) ``Phase diagram'', i.e., transition line
  between adsorbed and desorbed phases. Constructed from peak
  positions in~(a).}\vskip-.3\baselineskip
\label{fig5}
\end{figure}

\begin{figure*}[t]
\includegraphics{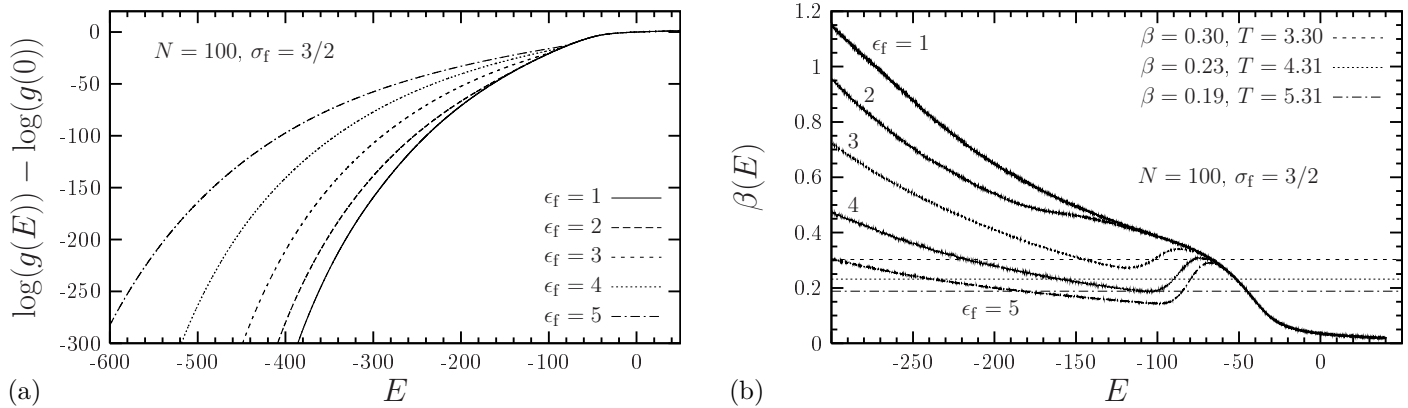}
\caption{(a) Logarithm of the density of states (proportional to the
  microcanonical entropy) for the system with $N=100$ monomers,
  $\sigma_{\mathrm{f}}=3/2$ and different string adsorption strengths
  $\epsilon_{\mathrm{f}}$. The convex intruder clearly emerges and
  becomes larger for increasing values of $\epsilon_{\mathrm{f}}$.
  (b) The derivatives of the functions in (a) with respect to $E$
  (proportional to the inverse microcanonical temperature). The lines
  mark the respective microcanonical adsorption temperatures obtained
  by the Maxwell construction in the backbending
  region.}
\label{fig6}
\end{figure*}

Finally, we short comment on thermodynamic properties of the
adsorption transition~\cite{tvlong10tbp}. For estimating the
finite-system transition temperature, we first identify peak positions
in the canonical specific-heat curves, associated to the
transition. In Fig.~\ref{fig5}(a) we plot these peaks for a~polymer
consisting of $N=100$ monomers and a~string with
$\sigma_{\mathrm{f}}=3/2$ for various values of the substrate adhesion
strength $\epsilon_{\mathrm{f}}$. In Fig.~\ref{fig5}(b), we plot the
transition temperatures corresponding to the peak positions depending
on $\epsilon_{\mathrm{f}}$. We find an almost linear increase of the
adsorption temperature with adsorption strength, a~behavior which was
in the same way observed in a~recent study, where this transition has
been studied for the same polymer model interacting with planar
substrates~\cite{monika09jpcb}.

A~more adequate description of the thermodynamic behavior of small
finite-size systems provides the micro\-canonical
analysis~\cite{gross01buch,christoph06prl,christoph08jcp,christoph09epl,taylor09jcp,monika10arxiv},
based on the fact that all the relevant information about the system
is encoded in its density of states $g(E)$. In Fig.~\ref{fig6}(a), we
plot the logarithm of this function, which is proportional to the
micro\-canonical entropy [$S(E)=k_\mathrm{B}\ln g(E)$]. The adsorption transition
is represented by the convex region in the micro\-canonical entropy. The
energetic width of this convex part is a~measure for the latent heat
which is nonzero for a~first-order-like transition. The derivative of
the micro\-canonical entropy with respect to the energy yields the
inverse micro\-canonical temperature $\beta(E)=dS(E)/dE$.  It is plotted
in Fig.~\ref{fig6}(b) and exhibits a~monotonic change in the
transition region (therefore called ``backbending effect''). The
microcanonical transition temperature is obtained by a~Maxwell
construction in that region, indicated by horizontal lines in
Fig.~\ref{fig6}(b).  We note, in particular, that for
$\epsilon_{\mathrm{f}}=1$ and $2$, backbending does not occur. The
inflection points in this energetic region indicate second-order-like
transitions and belong to the $\Theta$
transition in bulk. This agrees
with previous observations that the ``strength'' of the first-order
signal decreases with decreasing substrate adsorption
strength~\cite{monika10arxiv}. A more detailed analysis of the structural
transitions in the polymer--wire system by means of microcanonical
thermodynamics will be presented in a forthcoming paper~\cite{tvlong10tbp}.

\paragraph{Acknowledgment}
This work is supported by the Umbrella program under Grant No.~SIM6
and by supercomputer time provided by the Forschungszentrum J\"ulich under Pro\-ject
Nos.~jiff39 and jiff43.


\vspace{-6.6pt}

\begin{thebibliography}{10}

\bibitem{tv10prl}
Vogel, T. and Bachmann, M.,
Phys. Rev. Lett. {\bf 104} (2010) 198302.

\bibitem{bachmann10acie}
Bachmann, M., Goede, K., Beck-Sickinger, A.~G., Grundmann, M., Irb\"ack, A., and Janke, W.,
Angew. Chem. Int. Ed. {\bf 49} (2010) 9530.

\bibitem{gao03ea}
Gao, M., Dai, L., and Wallace, G.~G.,
Electroanalysis {\bf 15} (2003) 1089.

\bibitem{hasan09am}
Hasan, T., Sun, Z., Wang, F., Bonaccorso, F., Tan, P.~H., Rozhin, A.~G., Ferrari, A.~C.,
Adv. Mater. {\bf 21} (2009) 3874.

\bibitem{milchev01jcp}
Milchev, A. and Binder, K.,
J. Chem. Phys. {\bf 114} (2001) 8610.

\bibitem{bachmann05prl}
Bachmann, M. and Janke, W.,
Phys. Rev. Lett. {\bf 95} (2005) 058102.

\bibitem{bachmann06pre1}
Bachmann, M. and Janke, W.,
Phys. Rev. E {\bf 73} (2006) 020901(R).

\bibitem{monika09jpcb}
M\"{o}ddel, M., Bachmann, M., and Janke, W.,
J. Phys. Chem. B {\bf 113} (2009) 3314.

\bibitem{milchev02jcp}
Milchev, A. and Binder, K.,
J. Chem. Phys. {\bf 117} (2002) 6852.

\bibitem{srebnik07cpl}
Gurevitch, I. and Srebnik, S.,
Chem. Phys. Lett. {\bf 444} (2007) 96.

\bibitem{bergneuh91plb}
Berg, B.~A. and Neuhaus, T.,
Phys. Lett. B {\bf 267} (1991) 249.

\bibitem{bergneuh92prl}
Berg, B.~A. and Neuhaus, T.,
Phys. Rev. Lett. {\bf 68} (1992) 9.

\bibitem{wangl01prl}
Wang, F. and Landau, D.~P.,
Phys. Rev. Lett. {\bf 86} (2001) 2050.

\bibitem{stefan07prl}
Schnabel, S., Bachmann, M., and Janke, W.,
Phys. Rev. Lett. {\bf 98} (2007) 048103.\vadjust{\break}

\bibitem{tvlong10tbp}
Vogel, T. and Bachmann, M., preprint (2011).

\bibitem{janke08buch}
Janke, W., editor,
{\em Rugged Free Energy Landscapes}, volume 736 of {\em Lecture Notes
  in Physics},
Springer, Berlin, 2008.

\bibitem{janke98physA}
Janke, W.,
Physica A: Stat. Theor. Phys. {\bf 254} (1998) 164.

\bibitem{wenzel99prl}
Wenzel, W. and Hamacher, K.,
Phys. Rev. Lett. {\bf 82} (1999)\break 3003.

\bibitem{hansm99epjb}
Hansmann, U. H.~E.,
Eur. Phys. J. B {\bf 12} (1999) 607. 

\bibitem{hansm02prl}
Hansmann, U. H.~E. and Wille, L.~T.,
Phys. Rev. Lett. {\bf 88} (2002) 068105.

\bibitem{zhoubhatt05pre}
Zhou, C. and Bhatt, R.~N.,
Phys. Rev. E {\bf 72} (2005) 025701(R).

\bibitem{schnabel09cpl}
Schnabel, S., Vogel, T., Bachmann, M., and Janke, W.,
Chem. Phys. Lett. {\bf 476} (2009) 201.

\bibitem{schnabel09jcp}
Schnabel, S., Bachmann, M., and Janke, W.,
J. Chem. Phys. {\bf 131} (2009) 124904.

\bibitem{taylor09jcp}
Taylor, M.~P., Paul, W., and Binder, K.,
J. Chem. Phys. {\bf 131} (2009) 114907.

\bibitem{deutsch97jcp}
Deutsch, J.~M.,
J. Chem. Phys. {\bf 106} (1997) 8849.

\bibitem{reith10cpc}
Reith, D. and Virnau, P.,
Comp. Phys. Comm. {\bf 181} (2010) 800.

\bibitem{tv10procathens}
Vogel, T. and Bachmann, M.,
Phys. Proc. {\bf 4} (2010) 161.

\bibitem{tvmbtmja10tbp1} Vogel, T., Mutat, T., Adler, J., and Bachmann,
  M., \textit{Accurate modeling approach for the structural comparison
    between monolayer polymer tubes and single-walled nanotubes},
  Phys. Proc., in press (2011).

\bibitem{tvmbtmja10tbp2} Vogel, T., Mutat, T., Adler, J., and Bachmann,
  M., \textit{Morphological similarities of carbon nanotubes and
    polymers adsorbed on nanowires}, preprint (2011).

\bibitem{gross01buch}
Gross, D.~H.~E.,
{\em Microcanonical Thermodynamics},
World Scientific, Singapore, 2001.

\bibitem{christoph06prl}
Junghans, C., Bachmann, M., and Janke, W.,
Phys. Rev. Lett. {\bf 97} (2006) 218103.

\bibitem{christoph08jcp}
Junghans, C., Bachmann, M., and Janke, W.,
J. Chem. Phys. {\bf 128} (2008) 085103.

\bibitem{christoph09epl}
Junghans, C., Bachmann, M., and Janke, W.,
EPL (Europhys. Lett.) {\bf 87} (2009) 40002.

\bibitem{monika10arxiv} M\"{o}ddel, M., Bachmann, M., and Janke, W.,
  Phys. Chem. Chem. Phys. {\bf 12} (2010) 11548.

\end{thebibliography}

\end{document}